\documentstyle[12pt]{article}
\newskip\zatskip \zatskip=0pt plus0pt minus0pt
\def\matth{\mathsurround=0pt}

\def\atversim#1#2{\lower0.7ex\vbox{\baselineskip\zatskip\lineskip\zatskip
  \lineskiplimit 0pt\ialign{$\matth#1\hfil##\hfil$\crcr#2\crcr\sim\crcr}}}

\input epsf 

\begin{document}
 
\begin{titlepage}



\hspace*{\fill}\parbox[b]{3.4cm}{MSUHEP-70723 \\ July 1997\\ hep-ph/9707448\\
\vspace*{2.0cm}}


\vspace*{-0.5cm}

\begin{center}
\large\bf
$H\rightarrow ggg(gq\bar q)$ at Two Loops in the Large-$M_t$ Limit
\end{center}
 
\vspace*{0.3cm}
 
\begin{center}
Carl R. Schmidt \\
Department of Physics and Astronomy\\
Michigan State University\\
East Lansing, MI 48824, USA
\end{center}
 
\vspace*{0.3cm}
 
\begin{center}
\bf Abstract
\end{center}
 
\noindent

We present a calculation of the two-loop helicity 
amplitudes for the processes $H\rightarrow ggg$ and $H\rightarrow gq\bar q$ 
in the large-$M_t$ limit.  In this limit the calculation 
can be performed in terms of one-loop diagrams containing an effective 
$Hgg$ operator.  These amplitudes are required for the 
next-to-leading order (NLO) corrections to the Higgs transverse momentum
distribution and the next-to-next-to-leading order (NNLO) corrections to the
Higgs production cross section via the gluon fusion mechanism.
 
\end{titlepage}
 
\baselineskip=0.8cm
 
\section{Introduction}

The Higgs Boson $H$ is the only particle of the Standard Model remaining to 
be discovered.  Its role is to provide a simple mechanism to break
the electroweak gauge symmetry and to give mass to the weak gauge bosons and
the fermions.  Of course, it is possible that nature uses more than a 
single scalar boson for this purpose, but still
the Standard Model and its supersymmetric extension are the primary examples
of the class of symmetry-breaking models which interact weakly.
Therefore, the search for the Higgs boson is of the highest priority for
the Large Hadron Collider (LHC) at CERN.

The detection of the Higgs Boson above background at the LHC will be a
challenging task.  In particular, if the mass of the Higgs is below 
$\sim 140$ GeV, near the threshold for decay into $W$ boson pairs, the
detection of the Higgs is quite subtle.  Although the largest production
mechanism by far is gluon-gluon fusion, the equally large backgrounds require
the use of the $H\rightarrow\gamma\gamma$ decay channel, which has a branching
ratio of ${\cal O}(10^{-3})$.  To prepare for the search, we need the
best theoretical predictions possible, and this means the inclusion of
quantum chromodynamic (QCD) corrections to Higgs production and decay.  
Recent relevant reviews are given in reference \cite{spira}.

The Higgs production via gluon-gluon fusion proceeds at lowest order (LO)
through a quark loop.  This loop is dominated by the top quark, because
the Higgs coupling is proportional to the quark mass.  The two-loop 
next-to-leading order (NLO) QCD corrections have also been calculated
\cite{zerwas}, and they are quite large: $\sim 50-100\%$.  
An interesting feature of this NLO calculation is that it becomes much 
simpler in the limit of large top quark mass ($M_t\rightarrow\infty$).  
In this limit, one can integrate out the heavy top quark loop, leaving behind 
an effective gauge-invariant $Hgg$ coupling.  Thus, the number of loops at 
each order is reduced by one.  It has been shown that the NLO 
corrections in this large-$M_t$ limit give a good approximation to the
complete two-loop result over a large range of Higgs masses \cite{dawson}.
The large NLO correction suggests that even 
higher orders still may be important.  A soft-gluon resummation in the
large-$M_t$ limit has recently been performed, which gives an estimate
of the next-to-next-to-leading order (NNLO) corrections \cite{laenen}.  

Meanwhile, other groups have considered less inclusive quantities, 
such as the transverse momentum spectrum of the Higgs boson.  This 
observable has been considered at the one-loop Born level, both in the 
large-top-mass limit and with full $M_t$ dependence 
\cite{ptdist}. 
In addition, the effects of soft gluons have also been studied, which
modify the spectrum at small Higgs $p_{\perp}$ \cite{cprussel}.  However, 
a NLO calculation has not been done.  The real five-point $H\rightarrow gggg$ 
amplitudes which are needed have been calculated
by Dawson and Kauffman \cite{dandk} in the large-$M_t$ limit, and
recently Kauffman et al. \cite{kauffman} have calculated the five-point
amplitudes with light external quarks.  

In this paper we present the virtual corrections to the four-point 
$H\rightarrow ggg(gq\bar q)$ amplitudes in the large-$M_t$ limit, which 
completes the set of amplitudes needed to study the Higgs $p_\perp$ spectrum 
at NLO.  In this limit, the two-loop results can be computed from effective
one-loop diagrams.  The large-$M_t$ approximation to the Higgs $p_{\perp}$ 
spectrum will be good for some range of $M_H$ and Higgs $p_{\perp}$, and 
furthermore this calculation offers a check of the complete $M_t$-dependent 
result, when it should become available.  Moreover, these amplitudes
are necessary for a full NNLO calculation of the cross section in the 
large-$M_t$ limit \cite{kniehl}.

In addition to using the effective Higgs-gluon operator in the large-$M_t$ 
limit, we have also used several other techniques that have been found
 convenient in QCD loop calculations \cite{bdk}.  These include
the use of helicity spinors, color ordering, and background-field gauge.
In section 2 we discuss the details of the calculation, while in section
3 we present the amplitudes and discuss various cross-checks.
In section 4 we summarize our conclusions.

\section{Calculational Details}

In the large-$M_t$ limit the top quark can be removed from the
full theory, leaving a residual Higgs-gluon coupling term in the 
lagrangian of the effective theory:
\begin{equation}
{\cal L}_{\rm eff}\ =\ -{1\over4}\biggl[1-{\alpha_{s}\over3\pi}{H\over 
v}(1+\Delta)\biggr]\,{\rm Tr}\,G_{\mu\nu}G^{\mu\nu}\ .
\end{equation}
The finite ${\cal O}(\alpha_{s})$ correction to this effective operator 
has been calculated \cite{dawson} to be
\begin{equation}
\Delta\ =\ {11\alpha_{s}\over4\pi}\ .
	\label{opcorr}
\end{equation}
Following Mangano and Parke \cite{MP}, we use the unconventional 
normalization for the $SU(3)$ 
generator matrices ${\rm Tr}(T^{a}T^{b})=\delta^{ab}$ and $[T^a,T^b]=i\sqrt{2}
f^{abc}T^c$.  This will remove factors of $\sqrt{2}$ from the helicity 
amplitudes below.

As suggested by string theory methods \cite{bdk}, we use the 
background-field gauge to calculate the one-particle irreducible parts of the
Feynman diagrams.  The gluon field $G^\mu$ is split into a background
component $B^\mu$ and a quantum component $Q^\mu$, {\it i.e,}
 $G^\mu=B^\mu+Q^\mu$.
Two reasonable choices for the gauge-fixing term are
\begin{eqnarray}
{\cal L}_{\rm gf}^{(1)}&=&-{1\over2}\bigl(D^B_\mu Q^\mu\bigr)^2\ \nonumber\\
{\cal L}_{\rm gf}^{(2)}&=&-{1\over2}\biggl[1-{\alpha_{s}\over3\pi}{H\over 
v}(1+\Delta)\biggr]\bigl(D^B_\mu Q^\mu\bigr)^2\ 
\label{gaugefix}
\end{eqnarray}
where $D^B_\mu Q^\mu=\partial_\mu Q^\mu-(ig/\sqrt{2})[B_\mu\,,\,Q^\mu]$ is the
background-field covariant derivative of the quantum field.  The second
choice of gauge-fixing condition has the advantage that the $Hggg$ and the 
$Hgggg$ Feynman vertices retain
the same structure as the $ggg$ and $gggg$ vertices.  The simple organization 
of these vertices is the feature that makes the background-field gauge
the preferred gauge for doing one-loop calculations in QCD \cite{bdk}.  
However, the
price to be paid is that the Higgs boson now couples to the ghost fields.
We have verified that our results are the same using either gauge-fixing term.

We first consider the $H\rightarrow ggg$ amplitude.
For simplicity we take all particles to be outgoing so that we 
actually calculate the amplitude for the process $0\rightarrow H g_1g_2g_3$ 
with $p_H+p_1+p_2+p_3=0$.  The amplitude for 
gluons with helicities $\lambda_i$ and colors $a_i$ can be written
\begin{equation}
	{\cal M}\ =\ -\,{\alpha_s\over3\pi v}\,{g_s\over2}\,
        {\rm tr}\bigl(T^{a_{1}}[T^{a_{2}},T^{a_{3}}]\bigr)\,
	m(p_{1},\lambda_{1};p_{2},\lambda_{2};p_{3},\lambda_{3})\ .
\end{equation}
Note that the color ordering of the amplitudes is trivial here.

It is convenient to write these helicity amplitudes in 
terms of products of Weyl spinors $|p\pm\rangle$.  The polarization
vector for an outgoing gluon of momentum $p$ can be written \cite{xu}
\begin{equation}
\epsilon_{\pm}(p)^\mu\ =\ {\pm\langle p\pm|\gamma^\mu |q\pm\rangle
\over\sqrt{2}\langle q\mp| p\pm\rangle}\ .
\end{equation} 
The arbitrary reference vector $q$ satisfies $q^2=0$ and $q\cdot p\ne0$. 
A change in the reference vector just shifts $\epsilon(p)^\mu$ by 
a term proportional to $p^\mu$, which drops out of the gauge-invariant
helicity amplitude.  
There are two independent $Hggg$ helicity subamplitudes, which at tree 
level are
\begin{eqnarray}
m^0(1^+,2^+,3^+)&=&{-M_H^4\over\langle1\,2\rangle\langle2\,3\rangle
\langle3\,1\rangle}\nonumber\\
m^0(1^-,2^+,3^+)&=&{[2\,3]^4\over[1\,2][2\,3][3\,1]}\label{tree}\ .
\end{eqnarray}
Here we have used the notation $\langle ij\rangle =  \langle p_i-|p_j+\rangle
$ and $[ij]=  \langle p_i+|p_j-\rangle$.  These spinor products are 
antisymmetric and satisfy $\langle ij\rangle[ji] =2p_i\cdot p_j\equiv S_{ij}$.
All other subamplitudes can be obtained by invariance under cyclic 
permutations, charge conjugation, and parity.

We also consider the process $0\rightarrow H gq\bar q$
with $p_H+p+q+{\bar q}=0$.  The amplitude for a
gluon and quarks with helicities $\lambda, h, \bar h$ and colors 
$a, i, \bar \imath$ can be written
\begin{equation}
	{\cal M}\ =\ -\,{\alpha_s\over3\pi v}\,{g_s\over2}\,
        T^a_{i\bar \imath}\,m(p,\lambda;q,h;\bar q,\bar h)\ .
\end{equation}
At tree level we have
\begin{equation}
        m^0(g^+,q^-,\bar q^+)\ =\ {[p\bar q]^2\over[\bar q q]}\ .
\end{equation}
All other subamplitudes are either zero due to the requirements of helicity
conservation or can be related by charge conjugation and parity.

\section{The Helicity Amplitudes}

Figures 1 and 2 show representative box diagrams for the $Hggg$ and $Hgq\bar q$
amplitudes, respectively.  The Feynman diagrams have been evaluated with the
aid of the symbolic manipulation program MAPLE using the straightforward 
Passarino-Veltman reduction.  
For the sake of generality, we have regularized the loop integrals by 
continuing the loop momenta to $(4-2\epsilon)$ dimensions, while 
taking the number of helicity states of the internal gluons to be 
$(4-2\,\delta_R\,\epsilon)$.  Thus,
$\delta_R=1$ corresponds to the t'Hooft-Veltman scheme \cite{tv} and
$\delta_R=0$ corresponds to the four-dimensional-helicity scheme \cite{fdh}.

For the $Hggg$ amplitudes we obtain
\begin{eqnarray}
m^1(1^+,2^+,3^+)&=&m^0(1^+,2^+,3^+)\,{\alpha_{s}\over4\pi}
\,r_{\Gamma}\,\biggl({4\pi\mu^2\over-M_H^2}\biggr)^\epsilon
\,\Biggl[N_c U\nonumber\\
&&\qquad\qquad\qquad
+\,{1\over3}(N_c-n_f)\,{S_{31}S_{23}+S_{31}S_{12}+S_{12}S_{23}
\over M_H^4}\Biggr]\nonumber\\
m^1(1^-,2^+,3^+)&=&m^0(1^-,2^+,3^+)\,{\alpha_{s}\over4\pi}
\,r_{\Gamma}\,\biggl({4\pi\mu^2\over-M_H^2}\biggr)^\epsilon
\,\Biggl[N_c U\label{loop}\\ 
&&\qquad\qquad\qquad+\,{1\over3}(N_c-n_f)\,{S_{31}S_{12}
\over S_{23}^2}\Biggr]\ ,\nonumber
\end{eqnarray}
where the prefactor is
\begin{equation}
r_{\Gamma}\ =\ {\Gamma(1+\epsilon)\Gamma^{2}(1-\epsilon)
\over\Gamma(1-2\epsilon)}\ ,
\label{rgamma}
\end{equation}
and the universal singular factor is
\begin{eqnarray}
U&=&
{1\over\epsilon^2}\Biggl[
-\biggl({-M_H^2\over-S_{12}}\biggr)^\epsilon
-\biggl({-M_H^2\over-S_{23}}\biggr)^\epsilon
-\biggl({-M_H^2\over-S_{31}}\biggr)^\epsilon
\Biggl]\,+\,{\pi^2\over2}\nonumber\\ 
&&-\,\ln\biggl( {-S_{12}\over-M_{H}^{2}}\biggr)
\ln\biggl({-S_{23}\over-M_{H}^{2}}\biggr)
\,-\,\ln\biggl({-S_{12}\over-M_{H}^{2}}\biggr)
\ln\biggl( {-S_{31}\over-M_{H}^{2}}\biggr)
\, -\,\ln\biggl( {-S_{23}\over-M_{H}^{2}}\biggr)
\ln\biggl( {-S_{31}\over-M_{H}^{2}}\biggr)
  \nonumber\\
&&-\,2\,{\rm Li}_2\biggl(1-{S_{12}\over M_H^2}\biggr)
 \,-\,2\,{\rm Li}_2\biggl(1-{S_{23}\over M_H^2}\biggr)
 \, -\,2\,{\rm Li}_2\biggl(1-{S_{31}\over M_H^2}\biggr)\ .
 \label{Universal}
\end{eqnarray}

For the $Hgq\bar q$ amplitude we obtain%
\begin{equation}
m^1(g^+,q^-,\bar q^+)\ =\ m^0(g^+,q^-,\bar q^+)\,{\alpha_{s}\over4\pi}
\,r_{\Gamma}\,\biggl({4\pi\mu^2\over-M_H^2}\biggr)^\epsilon
\,\Bigl[N_c V_{1}+{1\over N_c}V_{2}+n_{f}V_{3}\Bigr]\ ,
\label{quarkloop}
\end{equation}
with
\begin{eqnarray}
V_{1}&=&
{1\over\epsilon^2}\Biggl[
-\biggl({-M_H^2\over-S_{gq}}\biggr)^\epsilon
-\biggl({-M_H^2\over-S_{g\bar q}}\biggr)^\epsilon
\Biggl]
\,+\,{13\over6\epsilon}\biggl({-M_H^2\over-S_{q\bar q}}\biggr)^\epsilon
\nonumber\\ 
&&-\,\ln\biggl( {-S_{gq}\over-M_{H}^{2}}\biggr)
\ln\biggl({-S_{q\bar q}\over-M_{H}^{2}}\biggr)
\,-\,\ln\biggl( {-S_{g\bar q}\over-M_{H}^{2}}\biggr)
\ln\biggl( {-S_{q\bar q}\over-M_{H}^{2}}\biggr)
  \nonumber\\
&&-\,2\,{\rm Li}_2\biggl(1-{S_{q\bar q}\over M_H^2}\biggr)
 \,-\,{\rm Li}_2\biggl(1-{S_{gq}\over M_H^2}\biggr)
 \, -\,{\rm Li}_2\biggl(1-{S_{g\bar q}\over M_H^2}\biggr)
 \\
 &&+\,{83\over18}\,-\,{\delta_{R}\over6}\,+\,{\pi^{2}\over3}\,-\,{1\over2}{
S_{q\bar q}\over S_{g\bar q}}\ ,
 \label{Universal1}\nonumber\\
V_{2}&=&
\biggl[{1\over\epsilon^2}+{3\over2\epsilon}\biggr]
\biggl({-M_H^2\over-S_{q\bar q}}\biggr)^\epsilon
\,+\,\ln\biggl( {-S_{gq}\over-M_{H}^{2}}\biggr)
\ln\biggl({-S_{g\bar q}\over-M_{H}^{2}}\biggr)
 \nonumber\\
 &&+\,{\rm Li}_2\biggl(1-{S_{gq}\over M_H^2}\biggr)
 \, +\,{\rm Li}_2\biggl(1-{S_{g\bar q}\over M_H^2}\biggr)
 \\
 &&+\,{7\over2}\,+\,{\delta_{R}\over2}\,-\,{\pi^{2}\over6}\,-\,{1\over2}{
S_{q\bar q}\over S_{g\bar q}}\ ,\nonumber\\
 \label{Universal2}
V_{3}&=&
-\,{2\over3\epsilon}
\biggl({-M_H^2\over-S_{q\bar q}}\biggr)^\epsilon
\,-\,{10\over9} \ .\label{Universal3}
\end{eqnarray}

In these expressions, $N_c=3$ is the number of colors, $n_f$ is the number 
of light fermions, and ${\rm Li}_2$ is the dilogarithm function. The 
amplitudes are written for $S_{ij}<0$ and $M_H^2<0$, but they can be 
analytically continued to the physical region by  
letting $-S_{ij}=-S_{ij}-i\eta$ and $-M_H^2=-M_H^2-i\eta$ for 
$\eta\rightarrow 0^+$.  Note that these amplitudes are 
ultraviolet-unrenormalized amplitudes.  Including the renormalization 
gives the modification
\begin{equation}
m^1\ \rightarrow\ m^1+(\Delta+3\delta_g)\,m^0\ ,
\end{equation}
where $\Delta$ is the finite renormalization of the effective $Hgg$ operator,
given in eq.~(\ref{opcorr}) and $\delta_g$ is the gauge-coupling counterterm.  
Using the $\overline{\rm MS}$ subtraction scheme, the counterterm is
\begin{equation}
\delta_g\ =\ -\,{1\over\epsilon}\,{\alpha_s\over4\pi}\,\Gamma(1+\epsilon)
\,(4\pi)^\epsilon\,\biggl[{11N_c\over6}-{n_f\over3}\biggr]\ .
\end{equation}

The amplitudes satisfy a number of consistency checks.  In addition
to the previously-mentioned variation of the gauge-fixing condition, we
have also shown that the amplitudes are invariant under different choices
of reference vectors for the gluon polarizations.  The poles in $\epsilon$
have been verified to cancel against the dipole subtraction term of
Catani and Seymour \cite{CS}, and the $\delta_R$-dependent terms appear with
the correct coefficients to relate the two different 
regularization schemes \cite{schemes}.  Finally, we have checked that the 
amplitudes obey the correct one-loop splitting formulae \cite{collinear}
in all of the singular collinear limits.

\section{Conclusions}

In this paper we have computed the two-loop corrections to the $H\rightarrow
ggg$ and $H\rightarrow gq\bar q$ helicity amplitudes in the large-$M_t$
limit.  The use of an effective $Hgg$ operator in one-loop Feynman diagrams, 
along with many of the standard techniques for calculating one-loop QCD
amplitudes, has reduced the complexity of this calculation immensely.  These
amplitudes complete the set needed to perform a NLO calculation
of the Higgs $p_\perp$ spectrum in the large-$M_t$ limit, which is currently
in progress \cite{chris}.  In addition, they are part of the set needed 
for a complete NNLO calculation of the total cross-section in this limit.

\noindent {\bf Acknowledgements}

We would like to thank Vittorio Del Duca, Lance Dixon, Chris Glosser, and
Jim Amundson for useful discussions.

\newpage

\begin{figure}
\vskip-3.0cm
\epsfysize=20cm
\centerline{\epsffile{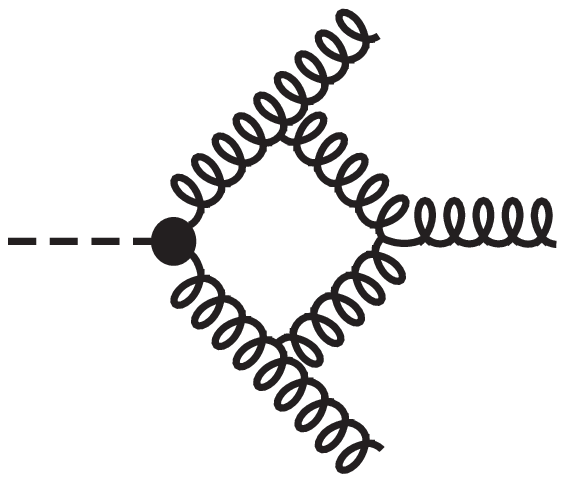}}
\vskip-11.0cm
\vskip-6pt
\baselineskip=12pt
Fig.~1: Box diagram for the $Hggg$ amplitude. The dot represents the
effective $Hgg$ vertex in the large-$M_t$ limit. 
\vskip2cm
\epsfysize=20cm
\centerline{\epsffile{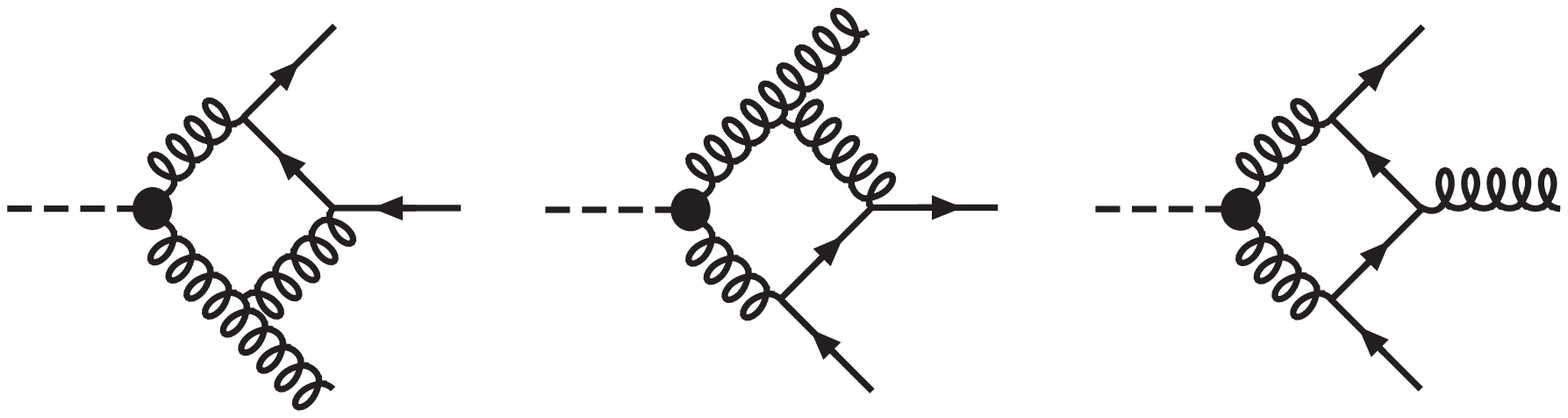}}
\vskip-11cm
\vskip6pt
\baselineskip=12pt
Fig.~2: Box diagrams for the $Hgq\bar q$ amplitude. The dot represents the
effective $Hgg$ vertex in the large-$M_t$ limit. 
\end{figure}

\end{document}